\documentclass[aps,prl,twocolumn,superscriptaddress,nofootinbib]{revtex4-2}

\usepackage{amsmath,amssymb,bm,braket}
\usepackage{graphicx}
\usepackage[percent]{overpic}
\usepackage{capt-of}
\usepackage[dvipsnames]{xcolor}
\usepackage{hyperref}
\usepackage{comment}

\hypersetup{
  colorlinks=true,
  linkcolor=black,
  citecolor=[rgb]{0.13,0.55,0.13},
  urlcolor=[rgb]{0.25,0.41,0.88}
}

\newcommand{\rvec}{\bm r}
\newcommand{\Sz}{\mathcal S_z}
\newcommand{\Sx}{\mathcal S_x}

\newcommand{\region}[1]{\raisebox{.5pt}{\textcircled{\raisebox{-.9pt} {#1}}}}

\begin{document}

\title{
Fuzzy Spectroscopy of Bound States in Massive Quantum Field Theories
}

\author{Joseph Taylor}
\affiliation{School of Physics and Astronomy, University of Leeds, Leeds LS2 9JT, United Kingdom}

\author{Matthew Yusuf}
\affiliation{School of Physics and Astronomy, University of Leeds, Leeds LS2 9JT, United Kingdom}

\author{Zlatko Papi\'c}
\affiliation{School of Physics and Astronomy, University of Leeds, Leeds LS2 9JT, United Kingdom}

\begin{abstract}

Mesons and glueballs are paradigmatic bound states of confining quantum field theories (QFTs), but their nonperturbative  spectroscopy in the continuum remains challenging beyond one spatial dimension. Here we perform such spectroscopy for the Ising QFT using a recently developed regularization based on noncommutative ``fuzzy'' geometry. On a thin fuzzy torus, we reproduce the universal low-lying $\mathbb E_8$ meson masses of the magnetically perturbed $(1{+}1)\mathrm{D}$ Ising QFT. By increasing the torus aspect ratio, we continuously track the second-lightest $\mathbb E_8$ meson as the system effectively crosses over from 1D to 2D. 
On the 2D fuzzy torus and sphere, we find a subthreshold scalar level and above-threshold response features consistent with previous estimates of glueball masses. 
The same excitations are revealed away from equilibrium using quench dynamics. Our results establish fuzzy geometries as nonperturbative spectroscopic probes of massive QFTs, including their bound-state evolution through dimensional crossover.
\end{abstract}

\maketitle

{\bf \em Introduction.---}Universal information about massive QFTs is encoded in their stable particles, bound states, and scattering continua. However, extracting this information nonperturbatively becomes especially difficult beyond one spatial dimension, where integrability is absent and finite-volume spectra become increasingly dense. The lowest multiparticle threshold is of crucial importance: an isolated pole below it can describe a stable particle, whereas above-threshold levels can mix with scattering states to form resonances of finite width. 

In the quantum Ising model, a longitudinal field confines domain-wall kinks into meson-like excitations~\cite{Sachdev11,mussardo2020statistical}. At the $(1{+}1)\mathrm{D}$ critical point, this magnetic deformation yields the celebrated integrable theory of eight stable particles whose mass ratios are governed by $\mathbb E_8$ exceptional Lie algebra~\cite{Zamolodchikov1989,perelomov2020remarks}. This spectrum has been observed experimentally~\cite{coldea2010quantum,Zhang2020E8,Zou2021E8}. By contrast, the same theory in $(2{+}1)\mathrm{D}$ is nonintegrable. Its lowest bound state is well established along several perturbation directions, whereas higher spectral features lie within the multiparticle continuum and require greater care~\cite{Caselle2000,Caselle2002,Nishiyama2008,nishiyama2010,Nishiyama2014,Nishiyama2016,caselle2002bound}. Along the thermal axis, Ising excitations are dual to glueballs of the confining $(2{+}1)\mathrm{D}$ $\mathbb Z_2$ gauge theory~\cite{Wegner1971,fradkin2021quantum}.

Such QFT spectra can be accessed directly in the continuum, i.e., without introducing a spatial lattice, if the model is defined on a noncommutative ``fuzzy'' geometry~\cite{Madore1992,DouglasNekrasovRMP}. A convenient practical implementation is based on embedding the QFT into a quantum Hall system, where projection to the lowest Landau level (LLL) naturally yields a finite Hilbert space without lattice discretization~\cite{Ippoliti:2018prb,Zhu23} (see also the recent review~\cite{FuzzyReview}). Importantly, translations or rotations of the underlying surface then remain exact even at finite size~\cite{Haldane:1983xm,Haldane85b}. The spherical version of this construction has recently enabled high-precision studies of many conformal field theories (CFTs) in $(2{+}1)\mathrm{D}$~\cite{Zhu23,Han24,voinea2024regularizing,Lauchli25,Ippoliti:2018prb,Zhou:2024qfi,chen24a,chen24b,Hofmann24,cuomo2024impurities,yang2025microscopic,fan2025simulatingnonunitaryyangleeconformal,cruz2025yangleequantumcriticalityvarious,miro2025flowingisingmodelfuzzy,Zhou24b,Hu2024,Zhou2024,Zhou2025,dedushenko2024isingbcftfuzzyhemisphere,He25,zhou2025chern,taylor2026conformal,voinea2026critical,zhou2026free,dey2026conformal,guo2026n,tang2026emergence,hao2026multi,huffman2026generalizing,meng2026quantum,eck20263d,janssens2026central,sarma2026fortuitous,dey2026wilson,feng2026studying,stergiou2026quantum}. 

\begin{figure}[t]
  \centering
  \includegraphics[width=\linewidth]{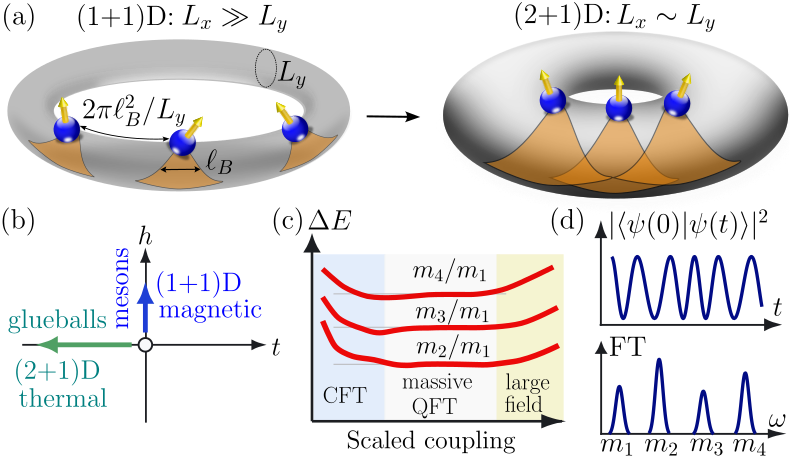}
  \caption{\label{fig:overview}
  \textbf{Fuzzy spectroscopy.}
(a) In the Landau gauge, the LLL orbitals are Gaussians whose width is set by the magnetic length $\ell_B$, while their centers are separated by
$2\pi\ell_B^2/L_y$. Varying $L_y$ tunes the overlap of these orbitals
continuously from an effectively $(1{+}1)\mathrm{D}$ thin-torus chain to
an isotropic $(2{+}1)\mathrm{D}$ geometry.
(b) Magnetic and thermal perturbations to the Ising fixed point generate meson and glueball excitations.
(c) Universal meson and glueball mass ratios emerge in a massive-QFT scaling window between the CFT-dominated and magnetic-field-dominated regimes.
(d) The same excitations can be detected in the quench dynamics as peaks in the Fourier spectrum of the return fidelity.
}
\end{figure}

At a conformal fixed point, the state--operator correspondence makes the fuzzy sphere especially powerful, allowing CFT data to be extracted directly from the energy spectrum~\cite{Cardy84,HuOPE2023,Han23,fardelli2025constructing,fan2024noteexplicitconstructionconformal,hu2025entropic,fardelli2026improving,dong2025numerical,wiese2026locating,yang2026conformal}. However, once a relevant perturbation opens a gap, the spectrum is no longer governed by this correspondence~\cite{Belin2018torus} and other fuzzy geometries, such as the torus, become equally natural. In particular, the torus admits a continuous interpolation between an effectively 1D thin-torus limit and an isotropic 2D system by varying the aspect ratio, Fig.~\ref{fig:overview}(a). This crossover has been insightful for understanding gapped quantum Hall phases~\cite{TaoThouless,Seidel2005,Bergholtz2005,Bergholtz2008,Nakamura2012,Nachtergaele2021,Mukherjee2023}, but has not previously been used to follow confined QFT excitations from 1D to 2D.

In this work, we use this geometric crossover to establish nonperturbative continuum spectroscopy of bound states in massive QFTs, summarized in Figs.~\ref{fig:overview}(b)-(d). With the Ising QFT as an illustrative example, we first calibrate the framework in the thin-torus limit, where we reproduce the exactly known $\mathbb E_8$ mass ratios of the $(1{+}1)\mathrm{D}$ Ising QFT. As the aspect ratio increases, we track the second-lightest $\mathbb E_8$ meson continuously to the isotropic 2D endpoint, where it remains below the two-particle threshold, while the next $\mathbb E_8$ branch enters the multiparticle continuum. The 2D fuzzy torus and sphere yield a subthreshold scalar level and higher response features near the previous Ising and glueball mass estimates. Finally, in End Matter, we demonstrate that the same excitations can be independently accessed in real-time dynamics following a quantum quench.

{\bf \em RG trajectories.---}Microscopic QFT regularizations must distinguish \emph{continuum} from \emph{infinite-volume} limit: increasing the number of LLL orbitals removes the ultraviolet (UV) cutoff and approaches the continuum, but infinite volume additionally requires the system to become large compared with the correlation length, $L/\xi\gg 1$. 

Near the Ising fixed point, the continuum action is
\begin{equation}
{\cal S}={\cal S}_{\rm CFT}
+t\int d^d x\,\epsilon(x)
+h\int d^d x\,\sigma(x),
\label{eq:action}
\end{equation}
where $\cal{S}_\mathrm{CFT}$ is the action at the Ising fixed point, and the thermal and magnetic field perturbations have renormalization group (RG) eigenvalues $y_t=d-\Delta_\epsilon$ and $y_h=d-\Delta_\sigma$~\cite{henkel2013conformal}. On a finite geometry of size $L$, the universal physics is controlled by the scaling variables $D=tL^{y_t}$ and $\mu=hL^{y_h}$. Holding $(D,\mu)$ fixed as $L$ increases defines the RG trajectories
\begin{equation}
t(L)=D L^{-y_t},\qquad h(L)=\mu L^{-y_h},
\label{eq:trajectory}
\end{equation}
which approach the fixed point as $L\to\infty$ with suppressed UV corrections. 

RG covariance implies the excitation masses are universal functions $F_i$ of a dimensionless coordinate:
\begin{equation}
m_i(t,h)=|h|^{1/y_h}
F_i\left(t/|h|^{y_t/y_h}\right).
\label{eq:massscaling}
\end{equation}
With Eq.~\eqref{eq:trajectory}, this gives
$Lm_i = |\mu|^{1/y_h} F_i(D/|\mu|^{y_t/y_h})$, which remains finite at fixed $(D,\mu)$ as $L\to\infty$. Since the correlation length $\xi\sim m_1^{-1}$ is set by the lightest mass, $L/\xi\sim Lm_1$ also remains finite at finite $\mu$, i.e., the UV cutoff is removed but the system remains only a few correlation lengths across~\cite{Cardy85,grimm1997scaling}. As we discuss below, the accessible system sizes in $(2{+}1)\mathrm{D}$ primarily probe this finite-volume regime, while in $(1{+}1)\mathrm{D}$ they can reach the infinite-volume limit $L/\xi\gg 1$.

{\bf \em Model.---}We realize the RG trajectories above using spinful electrons in the LLL at unit filling $\nu=N/N_\phi=1$, where $N$ is the number of electrons and $N_\phi$ is the number of  flux quanta. The electrons live on a torus of area
$L_xL_y=2\pi N$, with $\ell_B$ set to unity~\cite{Yoshioka83,Haldane85b,chakraborty2013quantum}. The shape of the torus is parametrized by its aspect ratio
$r=L_y/L_x$, with $r\ll1$ giving an effectively 1D thin
torus, while $r=1$ gives the 2D square torus.  

The two components form the
Ising pseudospin, described by the LLL-projected Hamiltonian
\begin{equation}
\begin{aligned}
H={}&2\!\int d^2r_1\,d^2r_2\,
V(\rvec_{12})
n_\uparrow(\rvec_1)n_\downarrow(\rvec_2)
- h_x\Sx - h_z\Sz,
\\
\mathcal{S}_\alpha\equiv{}&\int d^2r\, \Psi^\dagger(\rvec)\sigma^\alpha\Psi(\rvec), \quad \Psi^\dagger(\rvec)=\left(c_\uparrow^\dagger(\rvec),c_\downarrow^\dagger(\rvec)\right).
\end{aligned}
\label{eq:hamiltonian}
\end{equation}
The first term in $H$ is the Ising coupling, written in the second-quantized form in terms of annihilation operators $c_a(\rvec)$ for an electron with spin $a=\uparrow,\downarrow$ at position $\rvec$, with $n_a(\rvec)\equiv c_a(\rvec)^\dagger c_a(\rvec)$, $\rvec_{12}\equiv \rvec_1 - \rvec_2$, and $\sigma^\alpha$ are the standard Pauli matrices ($\alpha=x,y,z$). The fields  $h_x$ and $h_z$ encode the thermal and magnetic perturbations, respectively. The interaction $V$ is parametrized by the Haldane pseudopotentials $(V_0,V_1)=(4,1)$~\cite{Haldane:1983xm}.  In the absence of longitudinal field ($h_z=0$), the thin-torus limit has an Ising
transition at $h_x^c\simeq0.9175$~\cite{Han25}, while square-torus
finite-size scaling gives $h_x^c\simeq1.575$, see Supplementary Material (SM)~\cite{SOM} for further information.  

\begin{figure}[t]
  \centering
  \includegraphics[width=\linewidth]{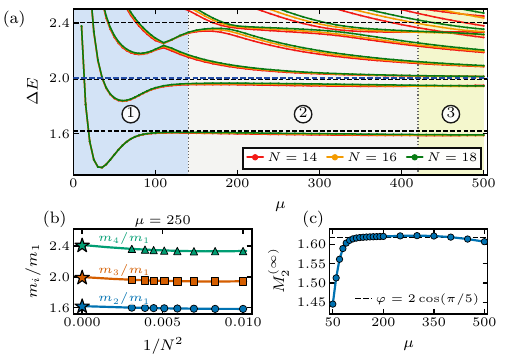}
  \caption{\label{fig:eight}
  \textbf{$(1+1)$D Ising benchmark.}
  (a) Thin-torus energy spectrum versus $\mu=h_zL_x^{15/8}$ for different $N$.
  Regions \region{1} and \region{3} are dominated, respectively, by residual CFT structure and microscopic large-field  corrections, while \region{2} is the universal massive scaling regime.  Horizontal dashed lines mark exact $\mathbb E_8$ ratios and
  the continuum threshold.   All gaps are divided by the lightest mass $m_1$.
  (b) $N\to\infty$ extrapolation of $m_2/m_1$, $m_3/m_1$, and
  $m_4/m_1$ at $\mu=250$.  At this field, the
  tracked $m_4$ branch is the \emph{third} energy-ranked level above $m_3$ in panel (a). Curves are quadratic least-squares fits in $1/N^2$, while stars mark
  the exact $\mathbb E_8$ values.  (c) Continuum estimate $M_2^{(\infty)}=\lim_{N\to\infty}m_2/m_1$as a function of $\mu$, showing the universal QFT window where $M_2^{(\infty)}$ matches the golden ratio
  $\varphi=2\cos(\pi/5)$.}
\end{figure}

\begin{figure*}[t]
  \centering
  \includegraphics[width=\textwidth]{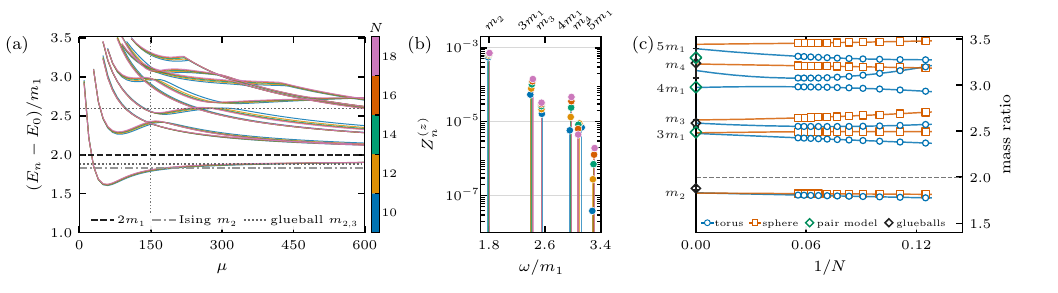}
  \caption{\label{fig:glueball-spectrum}
  \textbf{Continuum finite-volume $(2{+}1)\mathrm{D}$ Ising spectroscopy.}
(a) Square-torus spectra at $D=0$ for different $N$ (color bar) as a function of 
$\mu$.
The vertical line marks $\mu=150$.  Horizontal lines mark the continuum threshold (dashed), the $1.83m_1$ Ising estimate
(dash-dotted)~\cite{Caselle2000}, and the $1.88m_1$ and $2.594m_1$
glueball estimates (dotted)
~\cite{Caselle1998,Caselle2002}.
(b) The levels contributing to DSF, Eq.~(\ref{eq:dsf}), at $\mu=150$ for the same sizes as in (a), with 
stem heights given by $Z_n^{(z)}$.  Top labels denote the six bound states identified in (c).
(c) DSF-selected finite-volume gap ratios on the torus ($\mu=150$, open circles) and sphere ($\mu=26$, open squares).
All plotted sphere sequences have $L=0$; for the $4m_1$ torus level, we could only identify a nearby $L=1$ sphere counterpart, which is omitted in this plot~\cite{SOM}.
Lines are quadratic extrapolations in $1/N$, green diamonds are the
pair-model estimate (\ref{eq:composite}), and black diamonds are the glueball masses from Refs.~\cite{Caselle1998,Caselle2002}.  The
dashed line marks the continuum threshold.
}
\end{figure*}

{\bf \em $\boldsymbol{(1{+}1)\mathrm{D}}$ benchmark.---}We first demonstrate that our model (\ref{eq:hamiltonian}) reproduces the known bound-state Ising spectrum. 
On the thin torus, the rescaled field is $\mu=h_zL_x^{15/8}$ since $\Delta_\sigma=1/8$~\cite{onsager1944crystal,yang1952spontaneous,
henkel1989two}, where $L=L_x$ is the longer torus dimension.  We numerically diagonalize the Hamiltonian (\ref{eq:hamiltonian}) to obtain the low-lying energy spectrum $E_n$, identify the lightest gap $m_1=E_1-E_0$ and express all
other gaps in units of $m_1$.  Figure~\ref{fig:eight}(a) reveals three
regimes, cf. Fig.~\ref{fig:overview}(c).  At small $\mu$ (regime \region{1}), competition between finite-size CFT
behavior and massive-QFT scaling produces shallow minima in the mass
ratios~\cite{henkel1989two,henkel2013conformal}.  At large
$\mu$ (regime \region{3}), the longitudinal field is no longer weak and
short-distance corrections bend the levels away from their continuum values.
Between them (region \region{2}) lies the useful scaling window where several ratios become nearly
size independent and form broad plateaus in the vicinity of the exact $\mathbb E_8$ values.

To quantify this agreement, Fig.~\ref{fig:eight}(b)
extrapolates $m_2$, $m_3$, and $m_4$ at $\mu=250$ using
quadratic fits in $1/N^2$.  The $N\to\infty$ intercepts
lie within about $0.5\%$ of the exact $\mathbb E_8$ ratios
marked by the stars~\cite{Zamolodchikov1989}.  Figure~\ref{fig:eight}(c)
then applies the same extrapolation to demonstrate the robustness of $m_2/m_1$ across the wider range of $\mu$.  The continuum estimate is closest to the exact 
$\mathbb E_8$ value in the central scaling regime and departs from it
toward the CFT and large-$h_z$ regimes.  Previous tests of these
ratios can be found in
Refs.~\cite{Moore2011,Philipp22,grimm1997scaling, vovrosh2026mesonspectroscopyexoticsymmetries}.  Our agreement calibrates both the fuzzy regularization and the finite-size scaling prescription.  As
$N$ increases, the common scaling window broadens, while the large-$\mu$
turnover aligns at fixed bare $h_z$, confirming its microscopic
origin~\cite{SOM}.

{\bf \em $\boldsymbol{(2{+}1)\mathrm{D}}$ spectroscopy.---}On a square torus, we have $L=\sqrt{2\pi N}$ and $\mu=h_zL^{3-\Delta_\sigma}$, with $\Delta_\sigma=0.5181489$~\cite{Simmons-Duffin:2016gjk}. Taking the limit $N\to\infty$ at fixed $(D,\mu)$ removes the fuzzy UV regulator while keeping $m_1L$ finite, yielding 
continuum finite-volume levels, see 
Fig.~\ref{fig:glueball-spectrum}. The
large-$m_1L$ limit and the scaling window required to approach it are
analyzed in the SM~\cite{SOM}.

The massive spectrum depends on the RG direction away from the fixed point.  In
$(1{+}1)$D, the thermal deformation gives a free massive Majorana theory,
whereas the magnetic critical isotherm produces the interacting
$\mathbb E_8$ spectrum.  In $(2{+}1)$D, most high-precision glueball masses
have instead been obtained along the ordered thermal trajectory or mixed
thermal--magnetic crossover rays
~\cite{Caselle1998,Caselle2000,Caselle2002,Nishiyama2008,nishiyama2010,
Nishiyama2014,Nishiyama2016}.
Our Fig.~\ref{fig:glueball-spectrum}(a) follows the magnetic
critical isotherm, $D=0$, and varies $\mu$.  We select the optimal point
$\mu=150$ by minimizing the
slope of $m_2/m_1$ under a thermal perturbation~\cite{SOM}.

However, energy alone is insufficient to uniquely identify the relevant levels because of the dense spectrum, particularly above the two-particle threshold where the avoided crossings proliferate. We therefore resolve the spectrum with the aid of normalized dynamical structure factor (DSF) carrying zero momentum:
\begin{equation}
\begin{aligned}
S^{z}(\omega,\boldsymbol{q}=\boldsymbol{0})
&=\sum_{n>0} Z_n^{(z)}
\delta\left[\omega-(E_n-E_0)\right],\\
Z_n^{(z)}&=|\langle n|\Sz|0\rangle|^2/N^2.
\end{aligned}
\label{eq:dsf}
\end{equation}
Figure~\ref{fig:glueball-spectrum}(b) plots the weights $Z_n^{(z)}$ at $\mu=150$ for
the same system sizes as in panel (a).  Because both the ground state and
$\Sz$ transform in the fully symmetric $A_1$ representation of the square point group~\cite{SOM}, every displayed pole belongs to the $A_1$ sector.  Across the accessible $N$, the DSF
isolates a subthreshold scalar level at $m_2/m_1\simeq1.8$ together with higher DSF peaks above the continuum threshold. 

Figure~\ref{fig:glueball-spectrum}(c) follows the levels with nonzero DSF weight as $N\to\infty$. The extrapolations are consistent with previous glueball mass estimates in the literature, also indicated on the plot. As an independent check, we also perform extrapolations using the topologically-distinct fuzzy sphere geometry~\cite{SOM}. Both geometries consistently yield a subthreshold branch that corresponds to a stable particle and several levels above the threshold giving rise to DSF peaks. We refer to the latter as resonance candidates because an infinite-volume analysis is required to establish whether they correspond to stable masses or ultimately acquire a finite width.

In addition, we identify a few composite levels in Fig.~\ref{fig:glueball-spectrum}(b)-(c), denoted by $nm_1$. A heuristic pair-counting estimate gives
\begin{equation}\label{eq:composite}
 m_{nm_1}/m_1=n-\binom{n}{2}\Delta_2, \quad   \Delta_2=2-m_2/m_1\simeq0.17,
\end{equation}
which we call the ``pair model'' in Fig.~\ref{fig:glueball-spectrum}(c).
Table~\ref{tab:glueball-ratios} collects the bound-state gaps identified in Fig.~\ref{fig:glueball-spectrum}, along with their literature benchmarks and heuristic estimates where available, see also the SM~\cite{SOM}.  

Finally, End Matter complements the
static analysis above with real-time quench spectroscopy: in all three fuzzy geometries, the Fourier spectrum of the return fidelity recovers the
same frequencies identified by DSF, while the remaining additional peaks can be traced back to excited-state coherences, $m_i-m_j$.

\begin{table}[t]
\caption{\label{tab:glueball-ratios}
\textbf{Finite-volume bound state hierarchy}. All energies are in $m_1$ units. Torus and sphere values are our quadratic $1/N$ extrapolations for $N=8,\ldots,18$, with the uncertainty given in the parentheses. No $L=0$ sphere counterpart was identified for the torus $4m_1$ feature~\cite{SOM}. The glueball and lattice columns are previous results in the literature, with sources indicated in square brackets; ``pair'' refers to Eq.~(\ref{eq:composite}). }
\begin{ruledtabular}
\begin{tabular}{lccccc}
level  & torus & sphere & glueball & lattice & pair \\
\hline
$m_2$ & $1.832(1)$ & $1.829(2)$ & $1.88(2)$~\cite{Caselle1998,Caselle2002} & 1.83(3)~\cite{Caselle2000} & -- \\
$3m_1$  & $2.48(2)$  & $2.484(5)$ & --     & 2.45(10)~\cite{Caselle2000} & $2.49$ \\
$m_3$       & $2.589(9)$ & $2.623(2)$ & $2.59(4)$~\cite{Caselle1998,Caselle2002} & 2.55(12)~\cite{Nishiyama2016} & -- \\
$4m_1$ & $2.98(4)$  & -- & --      & $\approx 3$~\cite{Nishiyama2016} & $2.98$ \\
$m_4$   & $3.2(1)$   & $3.23(2)$ & $3.24(16)$~\cite{Caselle2002} & -- & -- \\
$5m_1$  & $3.40(2)$  & $3.45(3)$ & --         & -- & $3.30$ \\
\end{tabular}
\end{ruledtabular}
\end{table}

\begin{figure}[tb]
  \centering
  \includegraphics[width=\linewidth]{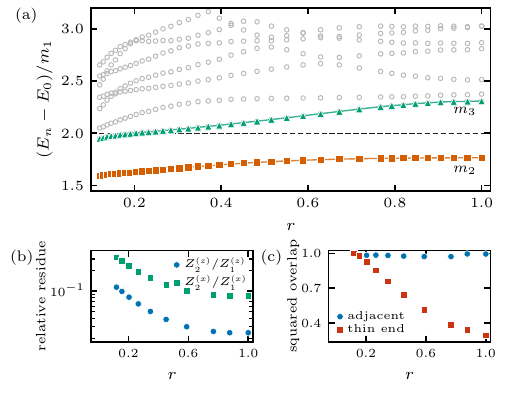}
  \caption{\label{fig:crossover}
  \textbf{Dimensional crossover of the QFT spectrum.}
  (a) Rescaled energy spectrum as a function of aspect ratio $r$ at fixed size $N=14$, $h_z=0.5$, and $h_x$ tuned to the finite-size critical line~\cite{SOM}. Filled symbols are overlap-tracked branches, open circles are all other levels, and the dashed line is the continuum threshold. The level $m_2$ interpolates smoothly between the 1D and 2D, while $m_3$ enters the continuum around $r\approx 0.2$. (b) DSF matrix element ratio for $m_1$ and $m_2$ levels. (c) Squared overlaps of the $m_2$ state at ratio $r$ with the adjacent state at $r+\Delta r$ or with the thin-limit state at the smallest $r$. The former shows adiabatic continuity, the latter illustrates the largely different states at the two end points.}
\end{figure}
{\bf \em Dimensional crossover.---}Having established the endpoints, we
now ask which part of the $\mathbb E_8$ spectrum survives as the second
spatial direction opens.  Figure~\ref{fig:crossover} follows the 
spectrum from the thin to the square torus at fixed $N$.  Because the magnetic scaling
exponent $y_h$ changes between the endpoints, no single renormalized magnetic
variable can be held fixed.  We instead fix the microscopic field $h_z=0.5$
and tune $h_x$ along the finite-size critical line, thereby defining a
well-specified path through the same Hamiltonian family. The critical field $h_x^c$ changes with the transverse circumference $L_y$ and  we determine it from the scaling of the gap between the even and odd spin-flip sectors, see SM~\cite{SOM}. 

Figure~\ref{fig:crossover}(a) shows the resulting 1D-to-2D spectral evolution which tracks the $\mathbb{E}_8$ branches $m_1$, $m_2$ and $m_3$ based on the overlap between eigenvectors at the neighboring values of the aspect ratio.  The tracked $m_2/m_1$ ratio rises from $1.596$ to
$1.766$ but remains below the continuum threshold throughout the crossover.  By contrast, the
next thin-torus branch $m_3$ crosses the two-particle threshold around $r\approx 0.2$.

Continuity in energy, however, does not guarantee that the branch remains
physically visible.  Figure~\ref{fig:crossover}(b) shows that the relative matrix elements
$Z_2^{(\alpha)}/Z_1^{(\alpha)}$, $\alpha=x,z$, remain nonzero across the
entire path.  On the other hand, local continuity does not imply an unchanged microscopic
wave function.  Figure~\ref{fig:crossover}(c) shows that squared overlaps between sampled aspect ratios remain above 97\%, whereas the overlap with the
original thin-torus state falls to $\sim 29\%$ at the square endpoint.  The
relative binding, $2-m_2/m_1$, simultaneously decreases from $0.404$ to
$0.234$~\cite{SOM}.  Thus the subthreshold branch remains continuous and
DSF-visible as the geometry crosses over from 1D to 2D, even though its microscopic wave function changes substantially.

{\bf \em Conclusions.---}We have demonstrated that fuzzy geometries provide more than a critical-point regulator: they allow massive excitations to be resolved along RG trajectories in different spatial geometries.  The exact
low-lying $\mathbb E_8$ mass ratios validate the regularization and finite-size scaling
prescription for the $(1{+}1)\mathrm{D}$ Ising QFT.  In $(2{+}1)\mathrm{D}$, torus and sphere extrapolations yield bound state levels near previous scalar-glueball estimates.  Tuning the effective spatial dimensionality shows that the lightest nontrivial excitation remains below the continuum threshold and retains nonzero spin residues as the second spatial direction opens up, while higher features represent resonance candidates as they lie above the threshold and undergo avoided crossings.  

Our approach is reminiscent of continuum methods based on dimensional
interpolation, such as the $d=4-\epsilon$ expansion~\cite{zinn2021quantum} and conformal bootstrap
at noninteger $d$~\cite{Cappelli2019}, but here the spacetime
dimension remains fixed and the interpolation is generated by varying the
system anisotropy. While our results in $(2{+}1)\mathrm{D}$ establish regulator-extrapolated levels at finite volume, a central unresolved problem is reaching the infinite-volume limit. This would reveal how the discrete multiparticle levels reorganize into scattering continua and determine
whether above-threshold structures survive as resonances with well-defined masses and widths. Furthermore, extending our spectroscopy to higher excited states could also probe confinement-induced nonthermal scar-like states in continuum QFTs~\cite{James2019,Delacretaz2023}.
The same framework can be applied to bound states generated by relevant
deformations of other CFTs, allowing confinement and massive spectra to be
compared across different universality classes. 

\emph{Note added.---}Related preliminary results for the magnetically deformed $(2{+}1)\mathrm{D}$ Ising QFT on the fuzzy sphere, including the lowest subthreshold bound state, were presented by Fardelli at the IHES workshop~\cite{FardelliTalk2025}.

\begin{acknowledgments}
{\bf \em Acknowledgments.--}We would like to thank \mbox{Cristian} Voinea and Wei Zhu for helpful discussions. This work was supported by the Leverhulme Trust Research Leadership Award
RL-2019-015 and EPSRC Grants EP/Z533634/1, UKRI4851, UKRI1337, and EP/W524372/1. Numerical simulations have used DiagHam~\cite{diagham} and FuzzifiED~\cite{FuzzifiED} libraries on the University of Leeds ARC4 and AIRE HPC facilities.
\end{acknowledgments}

\bibliography{refs}

\section*{End Matter}

\begin{figure}
  \includegraphics[width=\columnwidth]{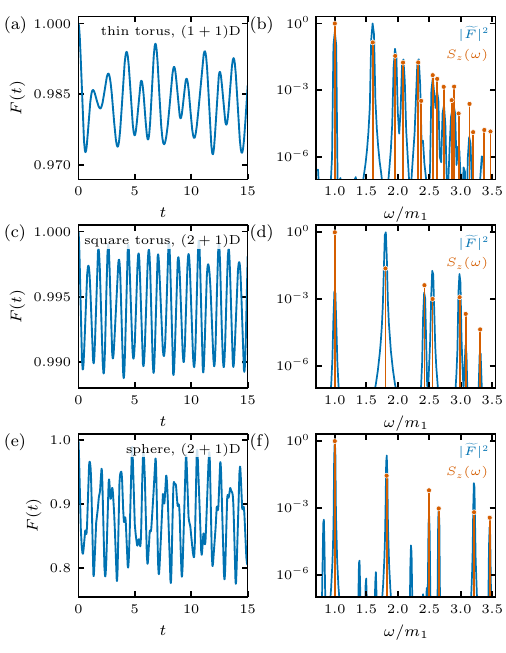}
  \caption{\label{fig:dynamics}
  \textbf{Dynamical signatures of bound states.}
  Each row shows the real-time return fidelity (left) and its Fourier spectrum (right). (a)-(b) The $(1+1)$D thin torus at
  $\mu=200$. (c)-(d) The $(2+1)$D square torus at $\mu=150$.  (e)-(f) The
  $(2+1)$D sphere at $D=0$, $\mu=26$.  Orange lines (b), (d), (f) are the matching
  equilibrium peaks of DSF.  Fidelity and DSF weights are
  normalized separately; frequencies are in units of the finite-volume gap
  $m_1$. All data are for $N=16$.}
\end{figure}

{\bf \em Quench spectroscopy.---}Figure~\ref{fig:dynamics}
complements the equilibrium analyses in Figs.~\ref{fig:eight} and
\ref{fig:glueball-spectrum} of the main text by asking whether the identified bound states are dynamically visible.  We consider a global quench protocol:
the ground state $|\psi(0)\rangle$ of the initial Hamiltonian is evolved under
the target Hamiltonian $H$, while monitoring the return fidelity
$F(t)=|\langle\psi(0)|e^{-iHt}|\psi(0)\rangle|^2$ (we set $\hbar=1$).  Expanding the initial state in the energy eigenbasis, the fidelity Fourier spectrum contains energy differences, $E_n-E_m$, between any populated eigenstates, rather than only the
gaps $E_n-E_0$, measured by the DSF.  Coincidence
with a DSF pole therefore cross-checks a finite-volume gap assignment.
Related quenches have previously been used to extract mass ratios in a
1D Ising spin chain~\cite{Moore2011}.

The top row of Fig.~\ref{fig:dynamics} benchmarks this protocol for the
$(1{+}1)\mathrm{D}$ Ising QFT on the thin torus. We prepare
the ground state at
$(h_x^{\rm in},h_z^{\rm in})=(0.7,0.05)$ and quench to $\mu=200$.
The Fourier peaks in panel (b) near $m_1$,
$m_2$, and the higher branches coincide with the
DSF peaks shown in orange, and they are seen to be in good agreement with the static spectrum in Fig.~\ref{fig:eight}(a).

The middle row applies the same protocol to the square torus.  We
prepare the ground state at
$(h_x^{\rm in},h_z^{\rm in})=(1.2,0.05)$ and quench to the magnetic working
point $h_x=1.575$, $\mu=150$.  Panel (d) compares the fidelity spectrum over
$0\leq t\leq100$ with the independently calculated DSF.
The principal peaks coincide with $m_1$, $m_2$, and the higher
operator-visible poles.  The additional feature near
$0.80m_1\simeq m_2-m_1$ is instead an excited-state coherence and therefore
need not appear in the DSF.

Finally,  the bottom row of Fig.~\ref{fig:dynamics} shows the corresponding quench on the fuzzy sphere.
The ordered-side ground state at
$(h_x^{\rm in},h_z^{\rm in})=(2.0,0.01)$ is evolved with the endpoint
Hamiltonian at $D=0$, $\mu=26$.  The stronger revivals in panel (e) reflect
appreciable occupation of several scalar eigenstates.  Panel (f) compares the
resulting fidelity frequencies with the equilibrium DSF in the
$L=0$ angular-momentum sector; the prominent peaks coincide with the
finite-volume scalar windows identified in
Fig.~\ref{fig:glueball-spectrum}.

In summary, quantum quench provides a dynamical visibility test of the finite-volume gaps identified through DSF and equilibrium spectral analysis, with consistent results across all types of fuzzy geometries, thus connecting the
static spectroscopy with real-time dynamics.

\clearpage
\onecolumngrid
\hypersetup{pageanchor=false}
\setcounter{equation}{0}
\setcounter{figure}{0}
\setcounter{table}{0}
\setcounter{section}{0}
\setcounter{page}{1}
\renewcommand{\theequation}{S\arabic{equation}}
\renewcommand{\thefigure}{S\arabic{figure}}
\renewcommand{\thetable}{S\arabic{table}}
\renewcommand{\thesection}{S\Roman{section}}
\renewcommand{\thepage}{S\arabic{page}}
\global\def\thepage{S\arabic{page}}
\makeatletter
\renewcommand{\theHequation}{S\arabic{equation}}
\renewcommand{\theHfigure}{S\arabic{figure}}
\renewcommand{\theHtable}{S\arabic{table}}
\renewcommand{\theHsection}{S\Roman{section}}
\makeatother

\begin{center}
{\large\bfseries Supplemental Material 
for\\[3pt]
``Fuzzy Spectroscopy of Bound States in Massive Quantum Field Theories''
}
\\[8pt]
Joseph Taylor, Matthew Yusuf, and Zlatko Papi\'c
\end{center}

\noindent
{\sl \small This Supplemental Material contains: (i) an introduction to the fuzzy-geometry
constructions; (ii) the finite-volume scaling conventions and a discussion of the approach to infinite volume; (iii) 
the thin-torus $\mathbb E_8$ benchmark and an overview of published $(2+1)$D mass estimates;
(iv) determination of the square-torus critical point; (v) the trajectory used to study the dimensional crossover; and (vi)
additional spectral diagnostics.
}

\section{Fuzzy regularization and microscopic model}
\label{sec:sm-fuzzy-model}

Fuzzy regularization, as used in this work, is based on realizing the desired QFT in a quantum Hall system projected to the lowest Landau level (LLL)~\cite{Zhu23}.  By construction, the LLL contains a finite number of orbitals determined by the number of magnetic flux quanta $N_\phi$ threading the system~\cite{Prange87}, which also sets the magnetic length $\ell_B$ as a fundamental short-distance cutoff. This naturally regularizes the theory without replacing the
continuous surface by a discrete lattice.  Increasing $N_\phi$ removes
the cutoff while preserving the spatial symmetries of the chosen geometry. Each electron in the LLL has an internal degree of freedom, labeled by
$a=\uparrow,\downarrow$, which will be used to encode an Ising pseudospin.  The charge sector is frozen by placing $N$ electrons at
integer filling factor $\nu=N/N_\phi=1$, while the pseudospin remains the relevant dynamical degree of freedom.

Let $\phi_m^{\mathcal G}(\mathbf r)$ be an LLL orbital on geometry
$\mathcal G$ and write the electron annihilation and spin operators as 
\begin{equation}
 c_a(\mathbf r)=\sum_m\phi_m^{\mathcal G}(\mathbf r)c_{ma},
 \qquad
 \mathcal S_\alpha=\sum_m
 \boldsymbol c_m^\dagger\sigma^\alpha\boldsymbol c_m , \quad \alpha = x,y,z,
 \label{eq:sm-lll-fields}
\end{equation}
where
$\boldsymbol c_m^\dagger=(c^\dagger_{m\uparrow},
c^\dagger_{m\downarrow})$.  All fuzzy geometries considered in this work use the same projected
Hamiltonian,
\begin{align}
 H_{\mathcal G}={}H_{\rm int}^{\mathcal G}
 -h_x\mathcal S_x-h_z\mathcal S_z,\quad
 H_{\rm int}^{\mathcal G}={}
 2\sum_{m_1,\ldots,m_4}
 \mathcal V^{\mathcal G}_{m_1m_2;m_3m_4}\,
c^\dagger_{m_1\uparrow}c^\dagger_{m_2\downarrow} c_{m_3\downarrow}c_{m_4\uparrow},
 \label{eq:sm-common-hamiltonian}
\end{align}
with the three terms in $H_{\mathcal G}$ encoding, respectively, the Ising coupling, transverse and longitudinal fields.
The only geometry dependence is in the LLL-projected interaction matrix element
\begin{align}
 \mathcal V^{\mathcal G}_{m_1m_2;m_3m_4}
 =\int_{\mathcal G}d^2r_1d^2r_2\,
 \phi_{m_1}^{\mathcal G*}(\mathbf r_1)
 \phi_{m_2}^{\mathcal G*}(\mathbf r_2)
 V(\mathbf r_1-\mathbf r_2) 
 \phi_{m_3}^{\mathcal G}(\mathbf r_2)
 \phi_{m_4}^{\mathcal G}(\mathbf r_1).
 \label{eq:sm-interaction-matrix}
\end{align}
The interaction  potential $V$ is specified by a short list of Haldane pseudopotentials~\cite{Haldane:1983xm}. 

The roles of the two fields in Eq.~\eqref{eq:sm-common-hamiltonian} are
distinct.  The transverse field $h_x$ flips the pseudospin and tunes the
thermal, or $\mathbb Z_2$-even, operator through the Ising transition.  At
small $h_x$, the system is ordered along $z$, whereas sufficiently large
$h_x$ produces a pseudospin paramagnet.  The longitudinal field $h_z$ couples
directly to the order parameter $\mathcal S_z$ and is the magnetic,
$\mathbb Z_2$-odd, perturbation.  At $h_z=0$, the global spin flip
$\mathcal P_x$, which exchanges $\uparrow$ and $\downarrow$, is an exact
$\mathbb Z_2$ symmetry and the states can be assigned even or odd parity.
A nonzero $h_z$ breaks this symmetry, selects one ordered vacuum, and confines
the domain walls.  Both fields are uniform and therefore preserve all orbital
symmetries of the underlying surface.  Total particle number is conserved, whereas $\mathcal S_z$ is not conserved when $h_x\neq0$.

\subsection{Torus}

For a rectangular torus, flux quantization gives
$L_xL_y=2\pi \ell_B^2 N_\phi$ , where the magnetic length $\ell_B$ will be set to unity.  At filling factor $\nu=1$ we take
$N=N_\phi$, leaving $N$ LLL orbitals labeled by a guiding-center index
$m\in 0,1,\ldots, N-1$.  We parameterize the geometry of the torus by its aspect ratio $r$, given by
\begin{equation}
 r=\frac{L_y}{L_x},\qquad
 L_x=\sqrt{\frac{2\pi N}{r}},\qquad
 L_y=\sqrt{2\pi Nr}.
 \label{eq:sm-torus-geometry}
\end{equation}
Our model is invariant under magnetic translations on the torus~\cite{Haldane85b}, which  enforces guiding-center momentum
conservation,
$m_1+m_2=m_3+m_4$ (modulo $N_\phi$).  Unless specified otherwise, we work in the zero-momentum sector. 
Orbital inversion is also exact.  For the square torus $r=1$, the spatial
symmetry is enlarged to the square point group; the ground state (and excited states of primary interest) belong to
its fully symmetric $A_1$ representation that we discuss in more detail below.  On the torus, we define the interaction using 
the Haldane pseudopotentials $(V_0,V_1)=(4,1)$ (regardless of the spin) and set higher pseudopotentials to zero. The relevant matrix elements can be found in the literature, e.g. Ref.~\cite{chakraborty2013quantum}.

\subsection{Thin torus or fuzzy circle}

The ``circle'' geometry used for $(1+1)$D is simply the thin-limit of the torus mentioned above:  we fix $L_y=3.25$ and take
$L_x=2\pi N/L_y$, so the torus becomes increasingly anisotropic as we grow $N$.  The LLL guiding centers then form an effectively
one-dimensional ring while retaining exact translation and inversion
symmetry.  The interaction matrix elements and fields are still those of
Eqs.~\eqref{eq:sm-common-hamiltonian} and
\eqref{eq:sm-interaction-matrix}.  For the above pseudopotentials, $(V_0,V_1)=(4,1)$ and $V_{\ell\geq2}=0$, the
$h_z=0$ Ising transition occurs at
$h_x^c\simeq0.9175$~\cite{Han25}.

\subsection{Sphere}

The sphere geometry has one important difference compared to the torus -- the topological quantum number called the Wen-Zee shift~\cite{WenZee}. The flux through the sphere is denoted by $2Q$ and the particle number is fixed to $2Q=N-1$.  The $2Q+1$ LLL orbitals are monopole harmonics with
$m=-Q,\ldots,Q$, and the sphere radius is $R^2=Q$~\cite{Wu:1976ge}.  The spherical
interaction matrix elements can be organized into pair angular-momentum
channels~\cite{Haldane83},
\begin{equation}
 \mathcal V^{\rm sph}_{m_1m_2;m_3m_4}
 =\sum_{\ell}V_\ell\sum_{M=-J}^{J}
 C^{JM}_{Q m_1,s m_2}
 C^{JM*}_{Q m_3,s m_4},
 \qquad J=2Q-\ell ,
 \label{eq:sm-sphere-matrix}
\end{equation}
where $C^{JM}_{Q m_1,Q m_2}$ is a Clebsch--Gordan coefficient and $V_\ell$
penalizes relative-angular-momentum channel $\ell$.  We take
$(V_0,V_1)=(4.75,1)$ and $V_{\ell\geq2}=0$, for which
$h_x^c\simeq3.16$ at $h_z=0$~\cite{Zhu23}.  The sphere preserves the full SO(3)
orbital rotation group, so eigenstates are labeled by total angular momentum
$L$. Our spectroscopy calculations use $L=0$, while the $L=1$ sector is discussed at the end of this SM.  

\section{Finite-volume scaling conventions}
\label{sec:sm-scaling}

Let $L$ denote the linear size appropriate to a given geometry $\mathcal G$.  We have $L=L_x$ on
the thin torus; $L=\sqrt{2\pi N}$ on the square torus; and
$L\propto R \sim \sqrt{N}$ on the sphere.  Close to the Ising fixed point, the
dimensionless thermal and magnetic coordinates are
\begin{equation}
 D=t_{\mathcal G}L^{y_t},\qquad
 \mu=h_{\mathcal G}L^{y_h},\qquad
 y_t=d-\Delta_\epsilon,\quad y_h=d-\Delta_\sigma .
 \label{eq:sm-scaling-fields}
\end{equation}
Similar expression can be written for general QFTs, with different perturbing operators and scaling dimensions. 

To leading order, the continuum thermal and magnetic fields in Eq.~(\ref{eq:sm-scaling-fields}) are related to
the microscopic fields by
$t_{\mathcal G}=\kappa_t^{\mathcal G}(h_x-h_x^c)$ and
$h_{\mathcal G}=\kappa_h^{\mathcal G}h_z$.  We absorb the nonuniversal metric
factors $\kappa_{t,h}^{\mathcal G}$ into the definitions of the scaling
coordinates and write 
$D=(h_x-h_x^c)L^{y_t}$ and $\mu=h_zL^{y_h}$. For a finite system, we can construct the dimensionless quantity
\begin{equation}
 {\cal E}_n(D,\mu;N)=L\,[E_n(D,\mu;N)-E_0(D,\mu;N)].
 \label{eq:sm-fss-spectrum}
\end{equation}
Taking $N\to\infty$ at fixed $(D,\mu)$ removes the
short-distance cutoff while retaining a finite value of $m_1L$, or equivalently $ {\cal E}_1$.  This 
defines a continuum \emph{finite-volume} scaling function.  In practice, for fixed $(D,\mu;N)$, we are interested in mass ratios ${\cal E}_n /{\cal E}_1 = (E_n-E_0)/(E_1-E_0)$ where $E_i$ are raw energies.  Therefore, we normalize the plotted gaps by the lowest nonzero gap $m_1 \equiv E_1 - E_0$.  An infinite-volume bulk mass ratio additionally requires a regime in which finite-volume corrections have saturated while the bare perturbations
remain small.  

\section{Thin-torus \texorpdfstring{$\mathbb E_8$}{E8} benchmark}
\label{sec:sm-e8}

For $L_y=3.25$ and $L_x=2\pi N/L_y$, the system is sufficiently anisotropic that, for $h_x=0.9175$ and $h_z=0$, it realizes the $(1{+}1)\mathrm{D}$ Ising critical point~\cite{Han25}.  Since $\Delta_\sigma=1/8$ in the $(1+1)$D Ising CFT, the
magnetic scaling variable used in Fig.~2 of the main text is $\mu=h_zL_x^{15/8}$.
The exact first four quasiparticle mass ratios are~\cite{Zamolodchikov1989}
\begin{equation}
\begin{aligned}
 \frac{m_1}{m_1}=1, \,\,\,
 \frac{m_2}{m_1}=2\cos\frac{\pi}{5}=1.6180\ldots, \,\,\,
 \frac{m_3}{m_1}=2\cos\frac{\pi}{30}=1.9890\ldots, \,\,\,
 \frac{m_4}{m_1}=4\cos\frac{\pi}{5}\cos\frac{7\pi}{30}
 =2.4049\ldots .
\end{aligned}
 \label{eq:sm-e8-ratios}
\end{equation}

At fixed $\mu$, the finite-size corrections to the gap ratio can be expressed as 
\begin{equation}
 M_a(N,\mu)=\frac{E_a-E_0}{E_1-E_0}
 =M_a^{(\infty)}(\mu)+
 \sum_{k=1}^{p}\frac{c_{a,k}(\mu)}{N^{2k}} .
 \label{eq:sm-e8-fit}
\end{equation}
In Fig.~2 of the main text, we assume $p=2$ and fit all available sizes $N$ to obtain the continuum values $M_a^{(\infty)}(\mu)$, see also Table~\ref{tab:sm-e8} for $a=2,3$. We choose the $\mu$ values in the intermediate plateau region of Fig.~2, as discussed below.

\begin{table}[]
    \centering
\begin{tabular}{ccc}
$\mu$ & $M_2^{(\infty)}$ & $M_3^{(\infty)}$\\
\hline
150 & 1.617637 & 1.977198\\
200 & 1.618540 & 1.982211\\
250 & 1.618616 & 1.979464\\
300 & 1.618192 & 1.975778\\
350 & 1.618039 & 1.984498\\
exact & 1.618034 & 1.989044
\end{tabular}
    \caption{Quartic $1/N^2$ extrapolations of the two lowest nontrivial
thin-torus ratios for representative $\mu$ values.  The last row gives the exact $\mathbb E_8$ values.}
    \label{tab:sm-e8}
\end{table}

The QFT scaling window has two competing boundaries discussed in the main text.  Dimensional analysis gives
\begin{equation}
 m_1L_x\sim\mu^{8/15},\qquad
 h_z=\mu L_x^{-15/8}.
 \label{eq:sm-e8-window}
\end{equation}
Increasing $\mu$ first suppresses infrared finite-volume corrections, but at
fixed $N$ it eventually makes the bare longitudinal field large and exposes
short-distance physics.  A useful window to observe universal QFT behavior is therefore
\begin{equation}
 m_1L_x\gg1 \quad \text{and} \quad h_z<1.   
\end{equation}
In the thin-torus limit, the above conditions are roughly satisfied for $140\leq\mu\leq360$, although at small values of $N$, the window narrows down for higher masses. At larger values of $\mu$ beyond this range, we observe that the spectrum fragments into multiples of the bare field $h_z$.

\begin{figure}[htb]
\centering
\includegraphics[width=0.52\textwidth]{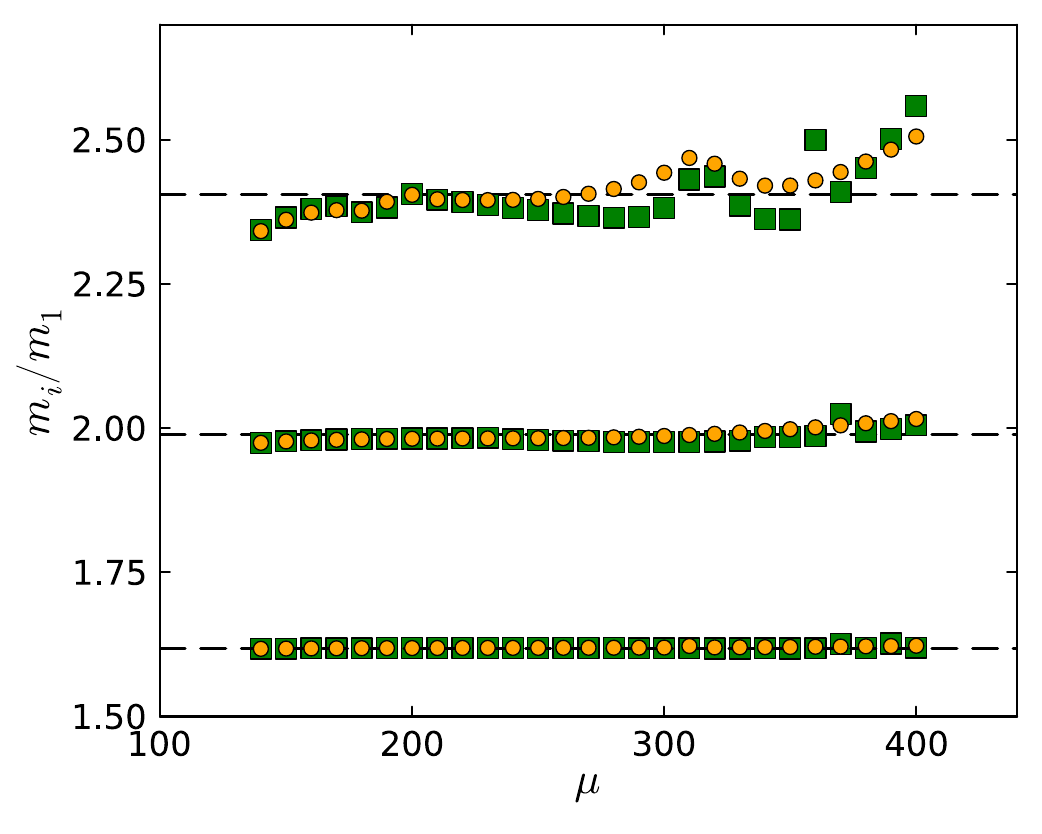}
\caption{\label{fig:sm-e8-all}
\textbf{Convergence of the lowest masses in $(1{+}1)\mathrm{D}$ towards their infinite-volume limit.} The mass ratios are plotted as a function of $\mu$.  Green squares use quartic fits in $1/N^2$
with $N=10,12,14,16,18$; orange circles use cubic fits with
$N=12,14,16,18$.  Dashed lines are the exact $\mathbb E_8$ ratios.  Agreement across the central window
and between the two extrapolations serves as an indication of the robustness of the results.}
\end{figure}

After taking the thermodynamic $N\to\infty$ limit, for an isolated stable particle, the leading remaining finite-volume correction is exponential~\cite{luscher1986volume},
\begin{equation}
 M_a^{(\infty)}(\mu)=\frac{m_a}{m_1}
 +A_a\exp\left(-c_a\mu^{8/15}\right)+\cdots .
 \label{eq:sm-e8-volume}
\end{equation}
Thus, a broad plateau as a function of $\mu$ (\emph{after} sending $N\to\infty$) can be used to estimate the intrinsic mass ratio in the \emph{infinite-volume} limit, assuming that the considered large $\mu$ values still keeps $h_z$ within the universal window.
Fig.~\ref{fig:sm-e8-all} shows that, in $\mathrm{(1+1)D}$, this happens for $m_2$ and $m_3$, and even for 
$m_4$ (with some uncertainty), which is the first state inside the two-particle continuum.

\section{Previous estimates of bound-state energies in \texorpdfstring{$(2+1)$D}{(2+1)D} Ising QFT}
\label{sec:sm-literature}

Table~\ref{tab:sm-literature} summarizes previous estimates of bound-state energies of the $\mathrm{(2+1)D}$ Ising QFT in the literature.  The quoted calculations employ different perturbation
directions, volumes, operators and extrapolation procedures, hence it is unsurprising that their predictions differ substantially beyond the lowest excitation. Nevertheless, different studies consistently point to the lowest ratio in the energy window $E\sim 1.8$--$1.9$ and a
higher scalar response near $E\sim 2.5$--$2.7$.  The glueball row is the literature benchmark used in Table~I of the main text.

\begin{table}[h]
\caption{\label{tab:sm-literature}
Selected published estimates of bound states in $(2+1)$D Ising QFT, including dual
theories.  A dash means that
the corresponding ratio was not quoted.}
\begin{ruledtabular}
\begin{tabular}{lccp{7.0cm}}
work & $m_2/m_1$ & $m_3/m_1$ & method or interpretation\\
\hline
Ref.~\cite{Caselle2000} & $1.83(3)$ & $2.45(10)$ & Monte Carlo Ising spectrum\\
Refs.~\cite{Caselle1998,Caselle2002} & $1.88(2)$ & $2.59(4)$ &
scalar $0^+$ glueballs of the dual $\mathbb Z_2$ gauge theory\\
Ref.~\cite{Caselle2002} & $1.828(3)$ & -- &
Bethe--Salpeter treatment of three-dimensional $\phi^4$ theory\\
Ref.~\cite{Nishiyama2008} & $1.84(3)$ & -- & finite-size lattice spectrum\\
Ref.~\cite{Nishiyama2014} & $1.84(1)$ & -- & finite-size lattice spectrum\\
Ref.~\cite{Nishiyama2016} & $1.81$ & $2.55(12)$ &
dynamical response; additional higher peaks were reported\\
Ref.~\cite{gao2025mesons} & $1.88792$ & $2.6568$ &
truncated coupled-chain construction\\
Ref.~\cite{gao2025mesons} & $1.94986$ & $2.80194$ &
$\mathrm{Ising}_h^2$ field theory\\
Ref.~\cite{Dusuel2010} & $1.81$ & -- & perturbative continuous unitary transformation\\
\end{tabular}
\end{ruledtabular}
\end{table}

The features labeled in columns by $nm_1$ ($n > 2$) in
Table~I of the main text are not additional fundamental excitations but denote composites of the fundamental $m_1$, bound pairwise. 
Writing the two-body binding as
$\Delta_2=2-m_2/m_1$ and recognizing that $m_2 \equiv 2m_1$ excitation is fundamental by construction, a simple pairwise model assigns an $n$-constituent
configuration the total binding $\binom n2\Delta_2$.  Neglecting higher
many-body corrections, we write
\begin{equation}
 m_{nm_1}=\left(n-\binom n2\Delta_2 \right)m_1 \,.
 \label{eq:sm-composite}
\end{equation}
With the pair binding energy $\Delta_2\simeq0.17$, this yields $2.49$, $2.98$, and $3.30$ for
$n=3,4,5$, in good agreement with our numerical data in Fig.~3 of the main text. 

\section{Square-torus critical point}
\label{sec:sm-square-critical}

\begin{figure}[t]
\centering
\includegraphics[width=0.65\textwidth]{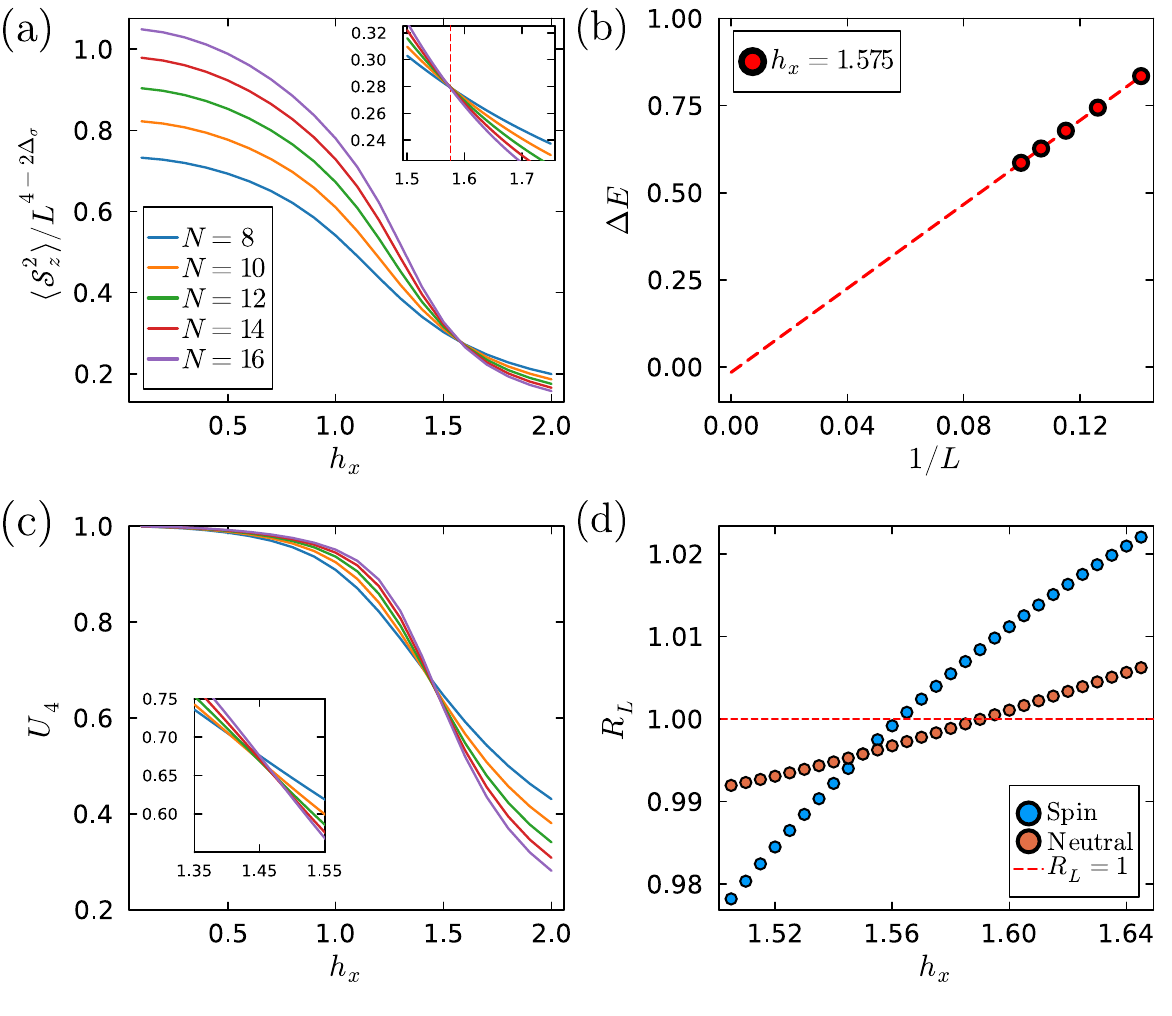}
\caption{\label{fig:sm-torus-fss}
\textbf{Determination of the $\boldsymbol{(2{+}1)\mathrm{D}}$ Ising critical point on the square torus.}
(a) Scaled order-parameter second moment for different $N$, with the inset enlarging  the crossing region.  The red dashed line indicates $h_x=1.575$.
(b) Energy gap at $h_x=1.575$ versus $1/L$. The linear extrapolation has intercept $-0.0151$.
(c) Binder ratio, with the crossing region enlarged in the inset.
(d) Consecutive-size ratios,  Eq.~\eqref{eq:sm-gap-ratio}, for the neutral and spin
gaps.  The combined diagnostics yield the critical field
$h_x^c=1.575$ used in the main text.}
\end{figure}

On the square torus, we have $L_x=L_y=L=\sqrt{2\pi N}$.  At $h_z=0$, we locate the
transition using standard finite-size diagnostics~\cite{Hamer_2000}. For example, the finite-size scaling of the Ising order parameter $\langle \mathcal S_z^2\rangle/L^{4-2\Delta_\sigma}$ reveals a crossing point at which the system becomes gapless, see Figs.~\ref{fig:sm-torus-fss}(a)-(b) respectively. We further examine the Binder ratio
\begin{equation}
 U_4=\frac{3}{2}
 \left(1-\frac{\langle\mathcal S_z^4\rangle}
 {3\langle\mathcal S_z^2\rangle^2}\right),
 \label{eq:sm-binder}
\end{equation}
which also shows a crossing near $h_x=1.575$ displayed in Fig.~\ref{fig:sm-torus-fss}(c).  Finally, for the raw energy gap $\Delta E_N$, we define the quantity
\begin{equation}
 R_L(h_x,N)=
 \frac{L_N\Delta E_N}{L_{N-2}\Delta E_{N-2}},
 \qquad L_N=\sqrt{2\pi N},
 \label{eq:sm-gap-ratio}
\end{equation}
which approaches unity at a $z=1$ critical point due to RG invariance of the numerator and denominator.  Both the neutral gap and the gap between the two spin-flip sectors pass through $R_L\simeq1$ near the
same field $h_x^c\simeq1.575$ used in the main text, albeit with stronger finite-size effect, see Fig.~\ref{fig:sm-torus-fss}(d).

\begin{figure}[t]
\centering
\includegraphics[width=0.85\textwidth]{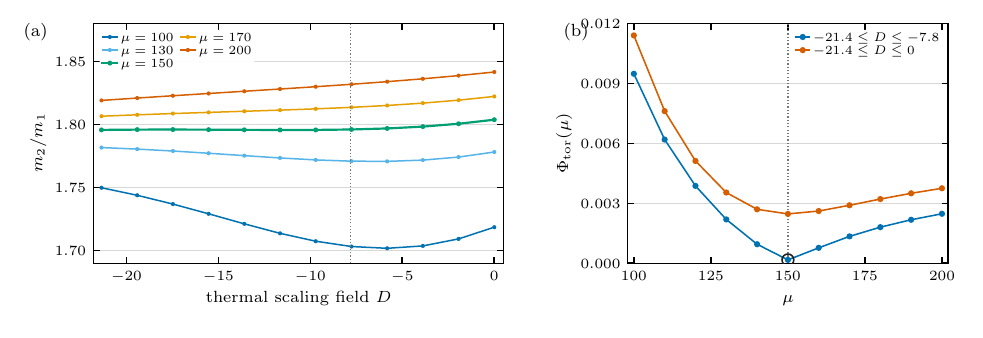}
\caption{\label{fig:sm-torus-flatness}
\textbf{Optimal magnetic perturbation $\mu$ for finite-volume analysis.}
(a) The lowest mass ratio $m_2/m_1$ at $N=16$ for several values of $\mu$ as a function of $D$.  The dotted line marks the upper end of the main
window (see text).  (b) Discrete slope norm, Eq.~\eqref{eq:sm-torus-flatness}, over
$-21.4\leq D\leq-7.8$ (blue) and over the wider interval
$-21.4\leq D\leq0$ (orange).  The circled minimum selects the optimal point $\mu=150$ used in the main text.}
\end{figure}

The chosen magnetic perturbation value $\mu=150$ used in the main text is fixed as follows.  At each candidate $\mu$ we follow
$m_2/m_1$ into the ordered phase and define the discrete slope norm
\begin{equation}
 \Phi_{\rm tor}(\mu)=
 \left[
 \sum_{D_i\in{\cal W}}
 \left[\nabla_D\!\left(\frac{m_2}{m_1}\right)_{D_i}\right]^2
 \right]^{1/2}.
 \label{eq:sm-torus-flatness}
\end{equation}
Here $\nabla_D$ is the discrete gradient on the sampled $D$ grid.  The main window $\cal W$ we focus on is $-21.4 \leq D \leq -7.8$, outside the near-critical region.  This window is equivalent to $1.025 \leq h_x \leq 1.375$ for $N=16$. Since $D=0$ corresponds to the critical field $h_x=h_x^c=1.575$, our choice of $D$-window therefore facilitates accessible sampling at uniformly spaced values of $h_x$ up to the critical isotherm $D=0$. 
 Fig.~\ref{fig:sm-torus-flatness}(a) shows the lowest mass ratio at $N=16$ over this window.  The slope changes sign and becomes nearly zero at $\mu=150$, as can be seen in panel (b) which evaluates Eq.~\eqref{eq:sm-torus-flatness} from the same data.  This criterion identifies the optimal point for our finite-volume calculations.

\section{Approach to infinite volume in \texorpdfstring{$(2+1)$D}{(2+1)D}}
\label{sec:sm-fig3-limits}

On the magnetic critical isotherm, RG covariance gives
\begin{equation}
 m_1L=C_1\mu^{1/y_h},\qquad
 h_z=\mu L^{-y_h}, \quad y_h=3-\Delta_\sigma,
 \label{eq:sm-two-limits}
\end{equation}
where the metric factor $C_1$ is nonuniversal. Increasing $N$ at
fixed $\mu$ thus sends the bare field to zero and removes the fuzzy UV
cutoff, but it still leaves $m_1L$ finite.  Since $m_1L$ is proportional to $L/\xi$, this means the system resides within a box of finite volume, with linear side spanning a few correlation lengths. This defines the continuum finite-volume limit
probed in Fig.~3 of the main text. More explicitly, panel (c) performs continuations of each ratio $R_a$ according to the form
\begin{equation}
 R_a(N,\mu)=R_a^{\rm cont}(\mu)
 +\frac{a_a(\mu)}{N}+\frac{b_a(\mu)}{N^2},
 \qquad \mu=150 ,
 \label{eq:sm-fig3-regulator-fit}
\end{equation}
Here, the superscript
``cont'' denotes removal of the regulator, not removal of the
spatial box. Table~I entries in the main text use quadratic fits to all available $N\geq 8$ data (torus at $\mu=150$; sphere at $\mu=26$). The uncertainties given in the parentheses are the differences between quadratic and cubic $1/N$ fits (also including the omission of a few lowest system sizes from the fitting). 

To determine the intrinsic value of the bulk ratio in \emph{infinite} volume would require the second limit
\begin{equation}
 \left(\frac{m_a}{m_1}\right)_{\rm bulk}
 =\lim_{\mu\to\infty}\lim_{N\to\infty}R_a(N,\mu),
 \label{eq:sm-fig3-bulk-limit}
\end{equation}
taken inside a window where $m_1L\gg1$ while the microscopic field
$h_z=\mu L^{-y_h}$ remains small.  The order matters: raising $\mu$ at fixed
$N$ eventually probes the microscopic large-field regime instead of improving
the continuum volume. 

Thus, in finite-size calculations there will be a finite (small or perhaps even non-existent) window in $\mu$ where we can access the scaling in Eq.~(\ref{eq:sm-fig3-bulk-limit}). To test whether we can access the regime in Eq.~(\ref{eq:sm-fig3-bulk-limit}),  we
follow the logarithmic running of the lowest nontrivial ratio,
\begin{equation}
 R_2(N,\mu)=\frac{E_2-E_0}{E_1-E_0},\qquad
 \beta_\mu(N,\mu)=\partial_{\ln\mu}R_2
 =\mu\,\partial_\mu R_2 .
 \label{eq:sm-running-flow}
\end{equation}
A bulk plateau requires $\beta_\mu\to0$ along a sequence for which
$\mu\to\infty$ and $h_z\to0$.  Small running by itself is not sufficient:
a finite system in the large-$h_z$ regime also has $\beta_\mu\to0$ as
$R_2\to2$.  We therefore restrict our attention to field strengths $h_z\leq0.6$ which are not too large to make the spectrum fragmented by too large $h_z$. 

\begin{center}
\centering
\includegraphics[width=0.95\textwidth]{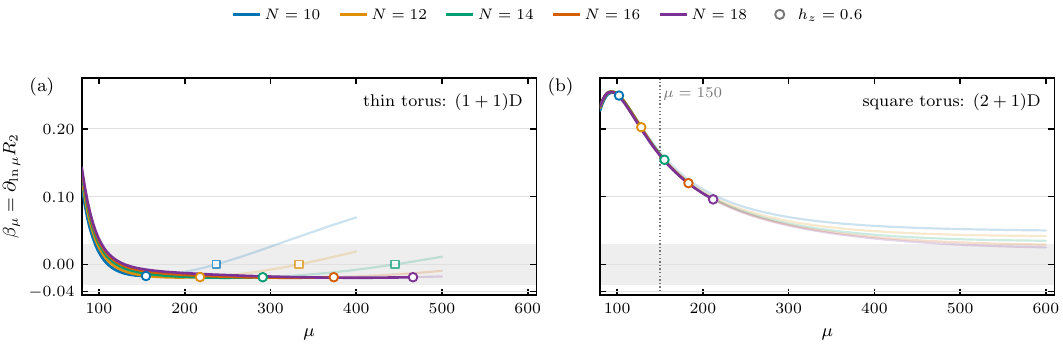}
\captionof{figure}{\label{fig:sm-infinite-volume}
\textbf{Running flow of the lowest mass ratio and approach to infinite volume.}
The shaded strip denotes $|\beta_\mu|\leq0.03$, i.e., the target region where the running coupling freezes.  Curves are plotted with solid lines while
$h_z\leq0.6$, open circles mark that cutoff, and subsequent data are faded. We use  cubic interpolation with the step $\Delta\mu=10$ to evaluate the derivative.
(a) On the thin torus, the data curves enter the target interval with the running coupling stuck close to zero.  Open squares mark the points where the running coupling starts to grow, indicating the departure from the universal QFT window due to large-$h_z$. The reliable range for extracting universal QFT data is the $\beta_\mu\approx 0$ plateau before the turning point. 
(b) On the square torus, the endpoints have positive
running.  At the previously chosen value $\mu=150$,
$\beta_\mu=0.161$--$0.166$ for $N=10$--$18$, and the subsequent approach to
the shaded strip is beyond the available system sizes. This suggests that our system-size range has not reached the infinite-volume limit. }
\end{center}

Figure~\ref{fig:sm-infinite-volume}(a) demonstrates that, in the $\mathrm{(1+1)D}$ case, the CFT and large-$h_z$ boundaries separate as $N$ grows. This leaves a robust  intermediate $\beta_\mu$ plateau, where universal QFT properties, such as the $\mathbb E_8$ mass ratios, can be reliably extracted. On the other hand, the isotropic torus results probing $\mathrm{(2+1)D}$ are different, Fig.~\ref{fig:sm-infinite-volume}(b).  Disregarding the very low fields dominated by the CFT, $\beta_\mu$ remains positive up to large $\mu$ values and for every
$N\leq 18$.  The $\beta_\mu$ values at $\mu=150$ vary by less than $0.005$
across those sizes but they are still not close to zero. Thus, finite-volume and microscopic-field effects have not become
parametrically separated in these system sizes.

\section{Trajectory for the dimensional crossover}
\label{sec:sm-cylinder}

To probe the dimensional crossover between the Ising QFT in $\mathrm{(1+1)D}$ vs. $\mathrm{(2+1)D}$, in the main text we varied the torus aspect ratio $r=L_y/L_x$ between the values $r\to 0$ (thin torus) and $r=1$ (isotropic torus). However, in doing this, the endpoint magnetic exponents differ, so there is no unique rescaled $\mu$ that can be held fixed as the aspect ratio
$r$ changes.  The crossover in Fig.~4 of the main text
therefore holds the microscopic value $h_z=0.5$ fixed to keep $\mu$ close to 150 for our considered system sizes, while the transverse field is determined independently at each $L_y$ assuming $h_z=0$. In other words, we follow the movement of the critical point with aspect ratio $r$ and look at the fixed $h_z$ perturbation applied to it.

For a fixed circumference $L_y$, increasing $N$ increases
$L_x=2\pi N/L_y$ and defines an effective 1D cylinder.  If
$E_0^\pm$ are the lowest zero-momentum energies in the two spin-flip sectors,
we evaluate the quantity $X_N$:
\begin{equation}
 X_N(h_x;L_y)=L_x\big(E_0^- -E_0^+\big)
 \label{eq:sm-prg}
\end{equation}
and locate $h_x^c(L_y)$ from the crossings of $X_N$ between successive system sizes.  We use a grid with minimum spacing 
$\Delta h_x=0.01$, yielding the critical-field trajectory summarized in Table~\ref{tab:sm-cylinder-line}. Note that 
the square ($r=1$) value  $h_x^c(L_y)=1.420$ is for the given finite $N$ and therefore it is not identical to value $1.575$ obtained after finite-size scaling that was used in Fig.~3 of the main text. 

\begin{table}[h]
\caption{\label{tab:sm-cylinder-line}
Selected critical fields for a few values of aspect ratio $r$ along the $N=14$, $h_z=0.5$ dimensional crossover path. Crossings were determined from sizes $(12,14)$ for $r \leq 0.819762$ and $(14,16)$ for larger $r$ (see text for details).}
\begin{ruledtabular}
\begin{tabular}{cc@{\qquad\qquad}cc}
$r$ & $h_x^c$ & $r$ & $h_x^c$\\
\hline
0.120077 & 0.920 & 0.628950 & 1.360\\
0.190912 & 1.110 & 0.819762 & 1.390\\
0.203987 & 1.130 & 0.847375 & 1.400\\
0.395601 & 1.280 & 1.000000 & 1.420\\
\end{tabular}
\end{ruledtabular}
\end{table}

At each point on the trajectory, we obtain 20 lowest zero-momentum eigenstates via exact diagonalization.  A few branches of interest are initialized at the thin endpoint and assigned at the next aspect ratio by maximizing squared
eigenvector overlap. For example, the two highlighted nontrivial branches in
Fig.~\ref{fig:crossover} evolve from $1.5956$ and $1.9505$ at
$r=0.120077$ to $1.7659$ and $2.3094$ at $r=1$.  The higher branch
passes the continuum threshold around $r\approx 0.2$.

The $m_2$ branch can be tracked using the matrix elements
\begin{equation}
 Z_n^{(\alpha)}=
 |\langle n|\mathcal S_\alpha|0\rangle|^2,
 \qquad \alpha=x,z .
 \label{eq:sm-residues}
\end{equation}
Figure~4 in the main text shows that the relative binding decreases from $0.404$ to $0.234$ as we sweep $r$, while the matrix elements drop as follows $Z_2^{(z)}/Z_1^{(z)} :0.115\to0.0245$ and $Z_2^{(x)}/Z_1^{(x)}:0.317\to0.0866$.
Furthermore, in Fig.~4 of the main text we compute the squared overlaps between the adjacent $r$ values, which lie between
$0.972$ and $0.995$.  On the other hand, the overlap of the $m_2$ state between the thin-torus and square torus limits is 
$0.295$, showing that the microscopic wave function changes substantially although the branch locally evolves continuously.

\section{Point-group symmetry of the square torus}
\label{sec:sm-symmetry}

\begin{figure}[htb]
\centering
\includegraphics[width=0.95\textwidth]{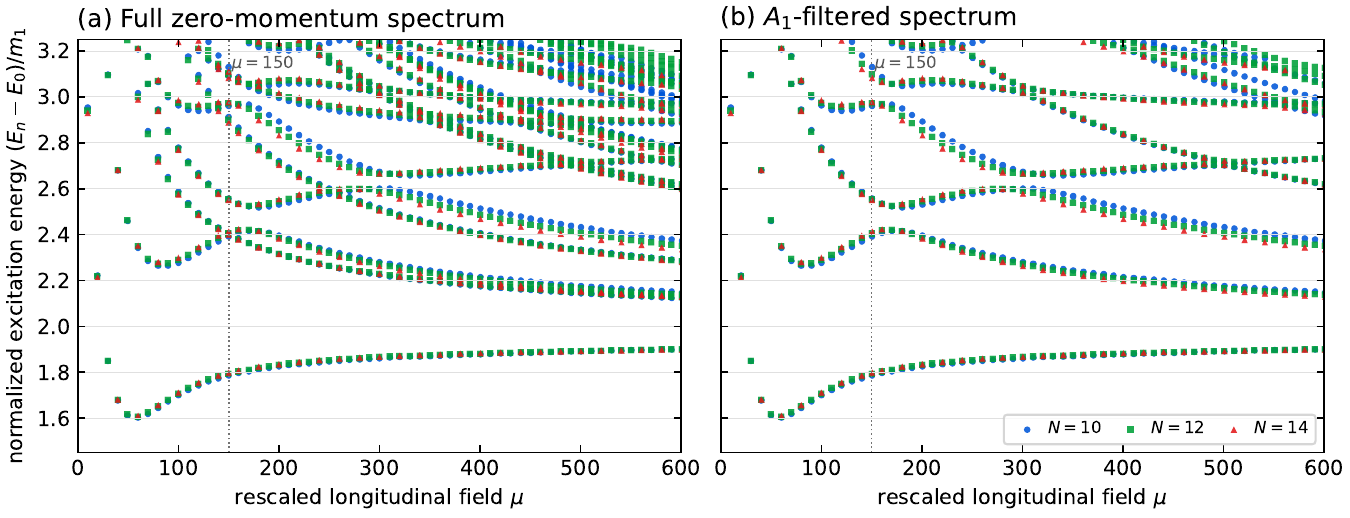}
\caption{\label{fig:sm-a1-spectrum}
\textbf{Point-group filtering of the square-torus spectrum.}
Zero-momentum spectra before (a) and after (b)
retaining levels visible to $A_1$ probe.  We remove levels that are visible only to the $B_1$ probe or which have zero weight to the $A_1$ probe.  This filtering shows that the apparent true crossings in (a)
occur between distinct symmetry sectors; the remaining $A_1$-visible branches
undergo avoided crossings, resulting in a simplified, physically-visible spectrum.}
\end{figure}

In the energy spectrum above the continuum threshold shown in Fig.~\ref{fig:sm-a1-spectrum}(a) [see also Fig.~3 of the main text], several levels appear to cross as $\mu$ is varied, while others undergo avoided crossings.  It is natural to ask whether these level crossings are intrinsic, i.e.,  occurring between states contributing to the same scalar response, or whether they can be explained by different symmetry of the levels.

At zero momentum, the ground state and $\mathcal S_x,\mathcal S_z$
operators transform in the fully symmetric $A_1$ representation of the
square point group. While this symmetry could be used to further block-diagonalize the Hamiltonian, for our purposes it is sufficient to resolve it \emph{a posteriori}, which can be done with the help of the following operators:
\begin{equation}
\begin{aligned}
 P_x(q)&=\mathcal S_z(-q_x)\mathcal S_z(q_x),&
 P_y(q)&=\mathcal S_z(-q_y)\mathcal S_z(q_y),\\
 P_{A_1}(q)&=\frac{P_x(q)+P_y(q)}{\sqrt2},&
 P_{B_1}(q)&=\frac{P_x(q)-P_y(q)}{\sqrt2}.
\end{aligned}
\label{eq:sm-pair-sources}
\end{equation}
By construction, $P_{A_1}$ is fully symmetric, while $P_{B_1}$ has $x^2-y^2$ character. 
We emphasize that $P_{A_1}$ is a convenient probe of $A_1$-symmetric response, not the full
projector onto the complete $A_1$ Hilbert-space sector.  The constituent spin  operators carry momentum and we will consider $q_k=2\pi k/L$, $k=1,2$.

Acting with one of these operators on the ground state produces the corresponding
probe state, $|v_\Gamma(k)\rangle=P_\Gamma(k)|0\rangle$, $\Gamma=A_1,B_1$.
The overlap $|\langle n|v_\Gamma(k)\rangle|^2$ measures how strongly level $n$ is
reached by a two-particle probe of symmetry $\Gamma$. Because Fig.~S5 concerns only levels above the continuum threshold, we remove the components
of $|v_\Gamma(k)\rangle$ along the already identified ground state, $m_1$,
and $m_2$ before normalizing these weights.  This does not change
$\langle n|P_\Gamma(k)|0\rangle$ for any above-threshold eigenstate, since
those eigenstates are orthogonal to the three removed states.

The ground state and the spin operators transform as $A_1$. 
The apparent true crossings in panel~(a) disappear under this filtering.
One of the removed branches can be associated with $B_1$ irreducible representation, while the remaining
removed branches are dark to the chosen $A_1$ probe.  The crossings therefore
do not occur between two branches of the operator-visible scalar response, but the levels instead
undergo avoided crossings accompanied by a transfer of spectral weight.
Figure~\ref{fig:sm-a1-spectrum} shows the filtered spectrum that only retains levels that transform according to $A_1$ irreducible representation. While the full spectrum [panel (a)] displays a mixture of level crossings and avoided crossing, the $A_1$-resolved spectrum [panel (b)] only contains avoided crossings, as expected for a non-integrable model. The
lowest $A_1$-visible levels  at $N=14,\mu=150$ are listed in Table~\ref{tab:sm-dsf-poles}.

\begin{table}[t]
\caption{\label{tab:sm-dsf-poles}
The lowest $A_1$-visible levels at $N=14$, $h_x=1.575$, and $\mu=150$.
$w_n^{(z)}=Z_n^{(z)}/\sum_{m>0}Z_m^{(z)}$ is the normalized residue of the
standard dynamical structure factor introduced in the main text.}
\begin{ruledtabular}
\begin{tabular}{ccc}
level assignment & $(E_n-E_0)/m_1$ & $w_n^{(z)}$\\
\hline
$m_1$ & 1.000000 & 0.973126\\
subthreshold $m_2$ & 1.799937 & 0.021565\\
lower response feature & 2.415576 & 0.003318\\
upper response feature & 2.552027 & 0.000849\\
near 3-body threshold feature & 2.979120 & 0.000779
\end{tabular}
\end{ruledtabular}
\end{table}

\begin{figure}[b]
\centering
\includegraphics[width=0.96\textwidth]{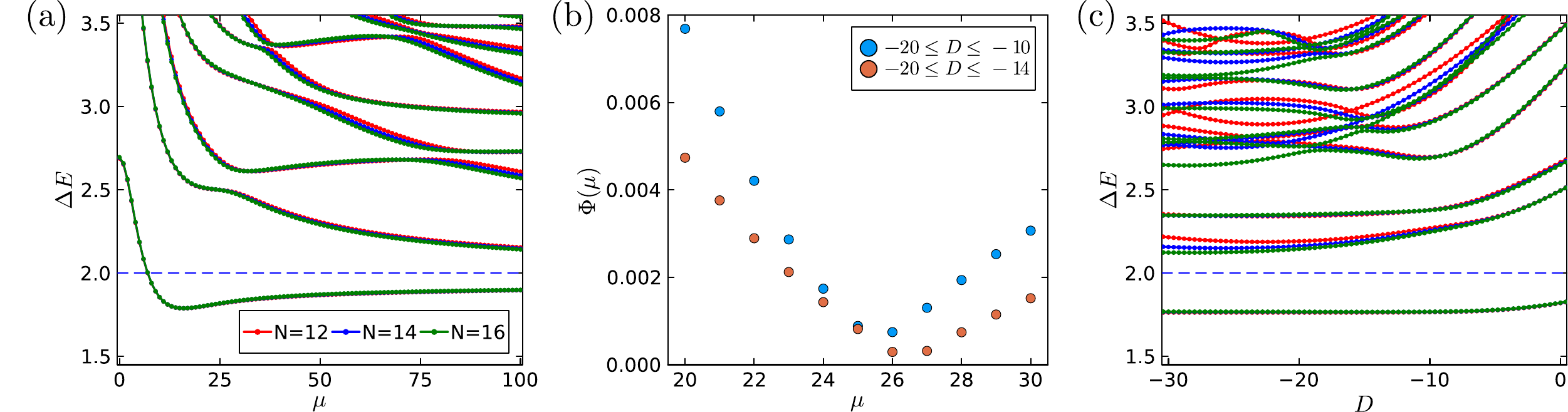}
\caption{\label{fig:sm-sphere-selection}
\textbf{Finding the optimal parameter point on the sphere.}
(a) Normalized spectrum at $D=0$ as a function of $\mu$.  (b) The slope norm 
$\Phi(\mu)$ [analog of Eq.~(\ref{eq:sm-torus-flatness})] for two choices of the fitting window.  (c) Spectrum as a function of $D$ at the optimal point $\mu=26$.  Red, blue, and green denote $N=12,14,16$.
The dashed line is the continuum threshold.}
\end{figure}

\section{Fuzzy sphere}
\label{sec:sm-sphere}

On the fuzzy sphere, the relevant length scale is the radius $R \sim \sqrt{N}$~\cite{Zhu23, Hu2024, hu2025entropic} and the scaling coordinates are
\begin{equation}
 D=(h_x-h_x^c) \sqrt{N} ^{3-\Delta_\epsilon},
 \qquad
 \mu=h_z \sqrt{N}^{3-\Delta_\sigma}.
 \label{eq:sm-sphere-fields}
\end{equation}
To identify the universal QFT window, we perform an analysis similar to that for the torus; see Fig.~\ref{fig:sm-sphere-selection}. For example, at fixed $\mu$, we track the spectrum as a function of $D$ and evaluate the spherical equivalent of the cost function in Eq.~\eqref{eq:sm-torus-flatness}. The minimum is found to occur at $\mu=26$, which is the value we used in the main text. 

One important difference between the sphere and torus is the strong proximity effect of the conformal fixed point at
$D=\mu=0$, where residual CFT effects strongly distort the spectrum, analogous to the behavior of
the $(1+1)$D Ising model discussed in the main text [Fig.~\ref{fig:eight}(a)].  Identifying the sphere's optimal scaled longitudinal field in Fig.~\ref{fig:sm-sphere-selection} therefore requires two conditions: not
only must $m_2/m_1$ exhibit a flat plateau in $D$, but the depth of this dip relative to the
critical point must also be minimized. This dip is indeed also minimized at our optimal choice of $\mu=26$, identified in Fig.~\ref{fig:sm-sphere-selection}. The same value of $\mu$ was used for spherical $N\to\infty$ extrapolations of the mass ratios in Fig.~3 of the main text based on Eq.~(\ref{eq:sm-fig3-regulator-fit}).

\subsection{\texorpdfstring{$L=1$ candidate for $4m_1$ bound state}{$L=1$ candidate for $4m_1$ bound state}}

\begin{figure}[h]
\centering
\includegraphics[width=0.6\textwidth]{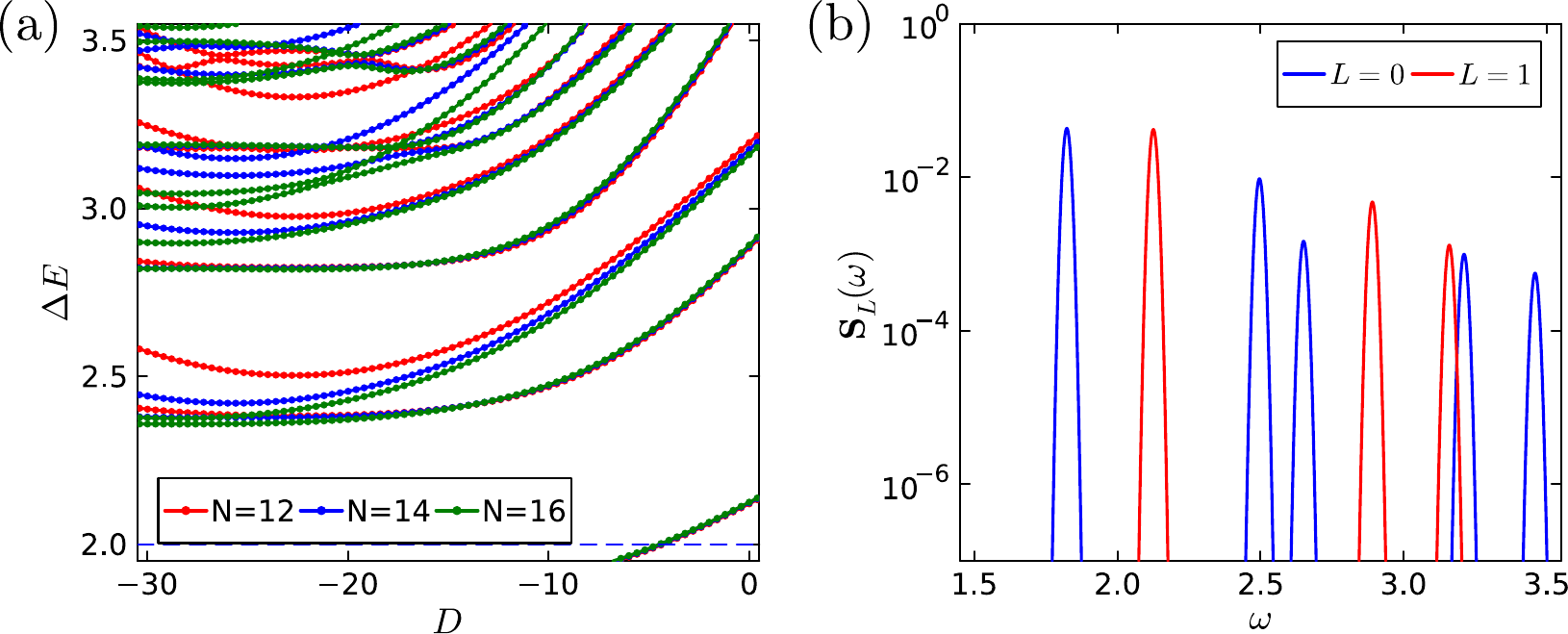}
\caption{\label{fig:sm-sphere-resolved}
\textbf{$L=1$ spectrum and DSF on the sphere.}
(a) Normalized $L=1$ spectrum versus $D$ at $\mu=26$ for $N=12,14,16$. (b) The spherical dynamical structure factor for $L=0$ (blue) and $L=1$ (red).  The $L=1$ peak lies near the $4m_1$ energy predicted by the pair model (\ref{eq:sm-composite}). In the DSF plot, the delta function is represented by a Gaussian of width
$\sigma=0.01$. 
}
\end{figure}

The torus $4m_1$ level does not appear to have a direct counterpart on the sphere in 
$L=0$ sector. However, the neighboring $L=1$ sector contains a level producing a  large contribution to DSF (suitably generalized to $L=1$), which lies in the expected energy range.
Figure~\ref{fig:sm-sphere-resolved} follows that sector into the ordered phase
and compares the spherical DSF for the $L=0$ vs. $L=1$ Fourier component of the spin operator. 
 The $L=1$ sequence extrapolates to $2.91(5)$, near the pair-model scale in Eq.~(\ref{eq:sm-composite}). 
 
\end{document}